\documentclass[singlecolumn,amsmath,amssymb,nofootinbib,showkeys]{revtex4-1}

\usepackage[TS1,T1]{fontenc}
\usepackage{graphicx}
\usepackage{dcolumn}
\usepackage{bm}
\usepackage{svg}
\usepackage[mathlines]{lineno}

\usepackage{subcaption}
\usepackage{booktabs}
\usepackage{easyReview}
\usepackage{siunitx}
\usepackage{float}
\usepackage{amsmath,bm}
\usepackage{widetable}

\usepackage{todonotes}
\usepackage{menukeys}

\usepackage{nicefrac}

\newcommand{\lambdabi}{\Lambda_\textrm{shuttle}}
\newcommand{\nubi}{\nu_\textrm{shuttle}}
\newcommand{\Dbi}{D_\textrm{shuttle}}
\newcommand{\etabi}{\eta}

\newcommand{\shuttlespeed}{v_0}

\newcommand{\emissions}{\mathcal{E}}
\newcommand{\quality}{\mathcal{Q}}
\newcommand{\trainEnergy}{e_\textrm{train}}
\newcommand{\shuttleEnergy}{e_\textrm{shuttle}}
\newcommand{\carEnergy}{e_\textrm{MIV}}

\newcommand{\fbi}{F}
\newcommand{\traffic}{\tilde \Gamma}

\newcommand{\funi}{(1-F)}

\newcommand{\dcut}{\bm{d_\textrm{c}}}

\newcommand{\dldc}{\langle d \rangle_{d < \dcut}}
\newcommand{\dgdctilde}{\langle \tilde{d} \rangle_{\tilde d > \tilde \dcut}}
\newcommand{\dldctilde}{\langle \tilde{d} \rangle_{\tilde d <  \tilde \dcut}}

\newcommand{\dshuttletilde}{\tilde D_{\mathrm{shuttle}}}
\newcommand{\deltamiv}{\Delta_\mathrm{MIV}}
\newcommand{\deltashuttle}{\Delta_\mathrm{shuttle}}
\newcommand{\deltashuttletilde}{\tilde\Delta_\mathrm{shuttle}}

\modulolinenumbers[5]

\begin{document}

\title{Bi-modal demand responsive transport in Berlin and Brandenburg.}

\author{Puneet Sharma}
 \email{puneet.sharma@ds.mpg.de}
\author{Stephan Herminghaus}
\author{Knut M. Heidemann}
\affiliation{Max-Planck Institute for Dynamics and Self-Organization (MPIDS), Am Fa\ss berg 17, 37077 G\"ottingen, Germany}

\date{\today}

\begin{abstract}
Bi-modal public transport system consisting of a rail-bound line service and a fleet of on-demand
shuttles providing connections to the line service stops is studied for the states of Berlin and
Brandenburg for user adoption of $x = 1\%$ and $x = 10\%$ of the total population.
\end{abstract}

\maketitle

\section{Introduction}
\label{intro}
\begin{figure*}
    \centering
    \includegraphics[width=0.8\textwidth]{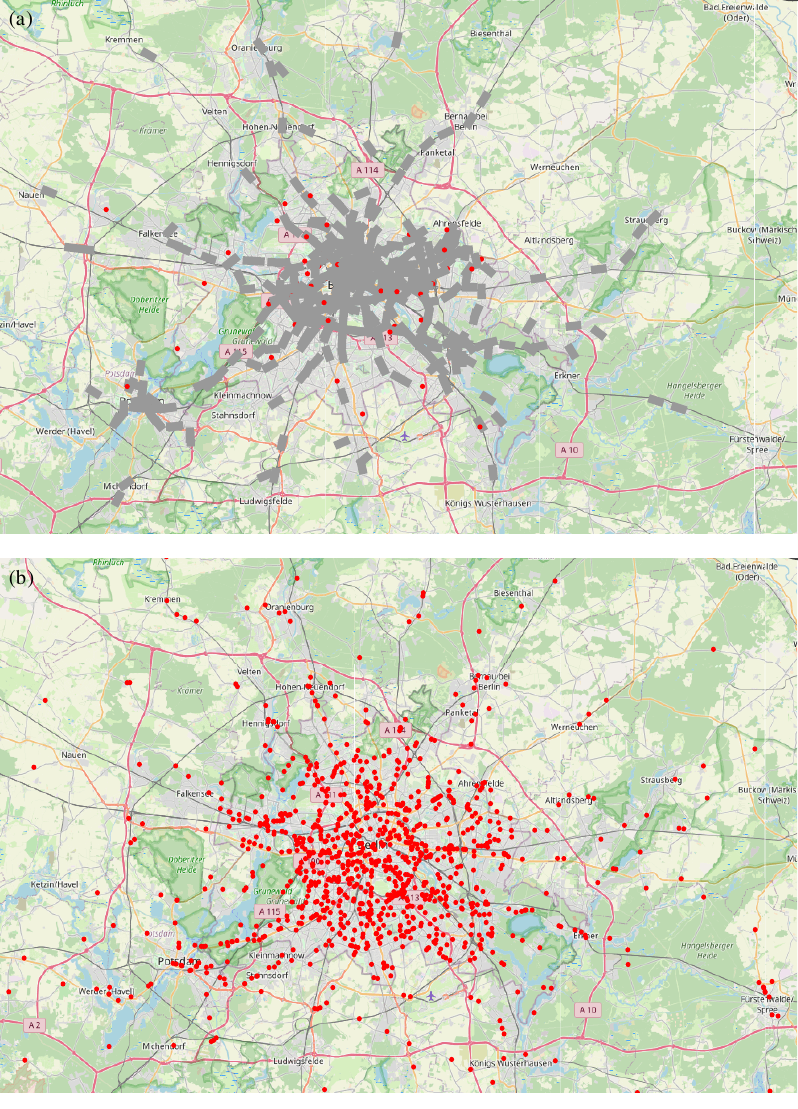}
     \caption{\textbf{Bi-modal transport network in Berlin and Brandenburg} A snapshot of simulations for $1\%$ user-adoption fraction $(x=0.01)$. \textbf{(a)} A bi-modal scenario where grey rectangles represent trains and red dots represent the shuttles. \textbf{(b)} MIV scenario where people use private cars to commute. Red dots represent private cars. The number of required shuttles in a bi-modal system (a) is much lower than the number of MIVS required in (b). See~Subsec.~\ref{subsec:traffic_results} for quantitative analysis.}
\end{figure*}

The energy-intensive nature of transportation, largely reliant on fossil fuels, has triggered a concerning surge in greenhouse gas (GHG) emissions. In the USA and Europe, this sector is a major contributor, accounting for more than a quarter of total GHG emissions from anthropogenic activities \citep{us_ghg}. The prevalent use of motorized individual vehicles (MIVs), i.e., private cars, for passenger transportation is a cause for concern. MIVs exhibit inefficiencies, as they involve the movement of over a ton of materials to transport a single person \citep{mackenzie2014determinants, tachet_scaling_2017}. This wastefulness has dire consequences, including air pollution and adverse environmental impacts \citep{air_emission_review,air_emissions_Kontovas2016,eeaair,ashok_pollution}, as well as contributing to traffic congestion \citep{traffic_singapore,traffic_Poland,traffic,traffic_greenhouse}.

Line services, represented by systems like light rails, have the potential to accommodate large numbers of passengers simultaneously, making them an ideal candidate for sustainable public transportation (PT) \citep{pitrzak_cities,ferbrache_cities,kato}. Despite their advantages, MIVs continue to dominate the global mobility market\citep{MIV_dominance, MIV_dominance_2} due to their perceived convenience\citep{kent2013secured}, flexible routes, and schedules.

Demand-responsive ride-pooling (DRRP) services present an alternative to both MIVs and line services. These services employ shuttle vehicles that adapt their routes and schedules based on user requests to combine individual user requests into an appropriate set of routes of the shuttle \citep{alonso-mora_-demand_2017}, thus providing door-to-door transportation. However, the trade-off between pooling and detour \citep{herminghaus_mean_2019,lobel_detours_2020} limits the achievable pooling efficiency of DRRP services \citep{zwick2021agent}, hindering their potential to reduce traffic and emissions significantly.

Previous studies \citep{bimodal_theory} have suggested a promising solution to this challenge, i.e., bi-modal transport. In a bi-modal public transport system, DRRP services are integrated with line services. Line services, characterized by fixed routes and schedules facilitate high vehicle occupancy and faster service. Meanwhile, DRRP shuttles offer seamless on-demand transportation to and from line service stops. Furthermore, shuttles are well-suited to serve short-distance routes with door-to-door service, where the use of line services is inefficient. This integration of line services and DRRP has been found to achieve high pooling efficiency for a simplified square grid geometry while ensuring user convenience. 

In a previous, mean-field-based approach, \cite{bimodal_theory} identified the key conflicting objectives of optimization as user convenience and energy consumption. It was found that the energy consumption may be reduced significantly relative to MIV, and traffic volume can be reduced by an order of magnitude relative to MIV. This suggested that bi-modal public transport systems have the potential to outperform customary public transportation as well as MIV in several respects. They also found that the overall performance of the system based on the objectives of emission and quality, does not vary dramatically with mesh size. Motivated by this, \citep{intermediate} studied the bi-modal transportation by means of agent-based simulation on a geometry akin to what one finds in real settings.  The study further reinforced the previous claim by \citep{bimodal_theory}.

While previous studies were motivating, they considered a spatially uniform demand with constant average request frequency. However, in realistic scenarios, demand is heterogeneous in space, due to non-homogeneous population density and individual mobility patterns, and fluctuating in time due to phenomena like rush hours. Furthermore, the trains were assumed to run periodically at the same constant speed throughout the system and the shuttles were also assumed to operate at a constant speed. These assumptions undermine the nuances in real settings. The network topology, for example, is much more complex in real settings and vehicles barely operate at the same constant speed, if at all. It is unclear how these heterogeneities affect the overall performance of the system. 

In previous studies, a prevalent assumption was that all transportation requests within a given study region exclusively relied on the public transit system, these requests were then assigned to bi-modal (shuttle-train-shuttle) or uni-modal (shuttles only) transportation. However, the reality is far more intricate, as user adoption of public transportation systems exhibits significant variability. In the present paper, we study the effects of such adoption variability on the holistic performance of bi-modal transportation systems. We explore two distinct scenarios: one where $x=1\%$ of the total population utilizes public transportation and another where the adoption rate increases to $x=10\%$ of the total population. For both scenarios, we assume that the rest of the people use private cars or MIVs to commute. Our study unveils the dynamics between user adoption patterns and the overall performance of bi-modal transportation systems, offering insights essential for optimizing their design and operation.

By studying bi-modal transportation in real cities, we aim to provide valuable insights into how effective bi-modal transportation can be when operated with the existing infrastructure of the rail network.

\section{Methods}
\subsection{Definition of the system}
For simulating a system of bi-modal transport, with on-demand shuttles and trains operating on lines, we deploy the open-source, multi-agent transport simulation framework MATSim \citep{MATSim}. We used the map from \cite{osm}. We used the passenger travel patterns that were artificially generated using the census data \citep{Berlin}. The data is provided by the Transport Systems Planning and Transport Telematics group of Technische Universität Berlin. We used an openly available General Transit Feed Specification (GTFS) dataset for the Berlin-Brandenburg region \citep{Berlin_transport} to generate MATSim public transport schedule and vehicle files and to add public transport links to the network. The dataset provides schedules and vehicles for various transport modes available, however, we only use rail-bound line services as a means of public transport. Each link on the transportation network has an associated speed for the vehicles.

\paragraph{Bi-modal transit system} The trains operating in the study area serve as the primary mode of transportation similar to what has been proposed in the previous study \cite{bimodal_theory}. These trains run according to the schedule described above.

The transit system is further characterized by a number of shuttles, $\mathcal{S}$, in the plane.
For the sake of conciseness and simplicity, we assume that the number of shuttles $\mathcal{S}$ is just sufficient to serve all user requests emanating in the system in a day.
Shuttles require energy $\shuttleEnergy$ per unit distance of travel. 
User requests served by DRRP/shuttles are subject to the constraint that the maximum accepted detour (traveled distance / direct distance) is $\delta_{m}=3$, the maximum waiting time is $\tau_{w,\,\text{max}}=5\, \unit{min} = 0.5 \cdot t_0$ and the maximum travel time is $\alpha\cdot t_\textrm{direct} + \gamma$, where $\alpha =3$, $\gamma=10\, \unit{min}$ are simulation parameters and $t_\textrm{direct}$ is the direct travel time. Note that we consider trains and shuttles a part of the public transit system.

\subsubsection{Parameters and objectives of operation}

\paragraph{Choosing the type of transport service}
\label{sec:choice}
  
A single user that adopts public transit in the model system may either be transported by uni-modal service, i.e., by shuttle (DRRP) only, or by bi-modal service, i.e., be brought from $\mathcal{P} = (x_{p},y_{p})$ to the nearest train station by means of a shuttle, followed by a train journey, which is again followed by a shuttle journey to $\mathcal{D}=(x_{d},y_{d})$. It is the task of the dispatcher system to decide, for each individual request $(\mathcal{P},\mathcal{D})$, whether the desired door-to-door service should be completed by uni-modal transportation (shuttles only) or bi-modal transportation (shuttle-train(s)-shuttle).
\cite{bimodal_theory} showed that the requested travel distance, $d=|\overline{\mathcal{PD}}|$, irrespective of the direction of travel, may serve as a reasonable discriminating parameter. Therefore, in order to choose the mode of transportation for an individual user request, we adhere to the previous policy \citep{bimodal_theory} of assigning user requests with travel distance $d>\dcut$ to bi-modal transportation (shuttle-train(s)-shuttle).  Shorter trips, i.e., user requests with travel distance $d \leq\dcut$, are assigned to uni-modal transportation (shuttle only).
The cut-off distance, $\dcut$, is the control parameter we will use to optimize the performance of the system.
It is in one-to-one correspondence to the fraction of bi-modal transportation $F(\dcut) = \int_{\dcut}^\infty p(y) \, dy$, out of the fraction of people, $x$, that adopt the public transit system, with $p(\cdot)$ the probability density of requested distances.

\paragraph{Service quality}
 
For the two $10\%$ and $1\%$ population scenarios, we define the service quality as the ratio between the average travel times by MIV and bi-modal transit, respectively, for the corresponding population fraction,
\begin{equation}
\label{eq:quality_word}
\quality = \frac{t_0}{\funi \cdot t_{\mathrm{uni}} + \fbi \cdot t_{\mathrm{bi}}}\,,
\end{equation}
where $\fbi = \fbi(\dcut)$ is the fraction of requests served by bi-modal transportation.

To compute the average travel time by MIV for a simulated scenario, we perform independent simulations where the MIV is the only allowed mode of transportation. If $t_{i}^\textrm{MIV}$ represents the travel time for user $i$, then $t_{0}$ can be obtained by averaging over all users in the system, i.e., $t_{0} = 1/\mathcal{N} \sum t_{i}^\textrm{MIV}$. Similarly, we compute the denominator in Eq.~\ref{eq:quality_word}, from  simulations by averaging over all users in the system.

In a mean-field approach, with square-grid geometry, \cite{bimodal_theory} derived an analytical expression for $\quality$ as
\begin{equation}
\label{eq:quality_math}
    \begin{aligned}
         \quality^{-1} = (1-\fbi)\cdot \underbrace{\left(\tilde{\tau}_{w} + \delta \dldctilde \right)}_{\tilde t_\text{uni}} + \fbi \cdot \underbrace{\left(  2\tilde{\tau}_{w} + 2\beta  \tilde{\ell}\delta + \frac{1}{\tilde{\mu}} 
         +  \frac{4}{\pi}\frac{\dgdctilde}{\tilde{v}_\textrm{train}}  \right)}_{\tilde t_\text{bi}}\,.
    \end{aligned}
\end{equation}
The $\ \tilde{ }\ $ indicates quantities non-dimensionalized via division by the respective unit ($D$, $t_0$, and $v_0$ as units for length, time, and velocity, respectively). $\dgdctilde$ is the mean of requested distances larger than $\dcut$ and $\delta$ is the average detour incurred by a user during the DRRP trip, $\tilde{\tau}_{w}$ is the average waiting time for shuttles and $\tilde \mu$ the train frequency, $\tilde{\ell}$ is grid constant or distance between two nearest train stations.
$\beta = \frac{1}{6}(\sqrt{2}+\log(1+\sqrt{2})) \approx 0.383$ is a geometrical constant, and $4\pi^{-1}\dgdctilde$ is the average distance traveled on trains.

\paragraph{Energy consumption}
\label{sec:energy_consumption}
 
We assess the overall energy consumption by transportation in the two use cases discussed above. To assess the overall energy consumption by the transportation system, we define the dimensionless energy consumption $\emissions$ as the ratio of total energy consumed when $10\%$ and $1\%$ population uses the bi-modal transportation and the total population using fleet of MIV, mathematically, $\emissions$ can be written as 
\begin{align}\label{eq:emission_word}
\mathcal{\emissions} \equiv \frac{\Delta_\mathrm{shuttle} \cdot \shuttleEnergy+\Delta_\mathrm{train} \cdot \trainEnergy }{\Delta_\mathrm{MIV} \cdot \carEnergy}\,,
\end{align}
where $\Delta_{\cdot}$ denotes the (mode-specific) total distance traveled in a unit cell of area $\ell^2$ per unit time.
$e_{\cdot}$ is the vehicle-specific energy consumption per unit distance, $x$  is the user-adoption fraction.
Note that $\mathcal E$ is already normalized with respect to the MIV energy consumption (denominator), as this is the door-to-door transportation system we intend to compare with. For $\emissions > 1$ ($<1$), energy requirement for bi-modal transportation is larger (smaller) than for MIV serving the same requests.

\citep{bimodal_theory} derived an analytical expression for a mean-field approach in a square grid geometry as follows. Both, uni-modal (shuttle only) and bi-modal trips, contribute to the total distance driven by shuttles per unit time due to requests from a unit cell of area $\ell^{2}$, hence
\begin{equation}\label{eq:d_shuttle}
    \Delta_\textrm{shuttle} = \frac{\nu E \ell^{2}}{\etabi} \left(\underbrace{\dldc\funi}_{\textrm{shuttle only}} + \underbrace{2\beta\ell\fbi}_{\textrm{two shuttle trips}}\right)\,,
\end{equation}
where $\eta$ is the DRRP pooling efficiency, which is the ratio of direct distance requested by the users and the  distance actually driven by the shuttles (for MIV, $\eta = 1$).
\cite{muehle2022} has observed that $\eta$ scales with demand $\Lambda$ roughly in an algebraic manner, $\eta(\Lambda)\approx\Lambda^{\gamma}$, with $\gamma\approx0.12$. In a bi-modal system, however, some of the demand $\Lambda$ is directed towards trains. Therefore, we need to compute an adjusted demand, $\lambdabi \equiv (E\nubi \Dbi^3)/\shuttlespeed$, considering shuttle trips only. $\nubi$ is the effective request frequency for shuttle trips and $\Dbi$ is the average distance of a shuttle trip.
One can show that \citep{bimodal_theory}
\begin{align}\label{eq:lmbd_shuttle}
  \Lambda_\mathrm{shuttle}  = \Lambda \, (1 + \fbi)^{-2}(\funi \dldctilde + 2 \beta \tilde{\ell} \fbi)^3\,.
\end{align}
We compute the theoretical pooling efficiency, $\eta$, according to the power law mentioned above 
\begin{align} 
\label{eq:eta_bi}
    \eta \equiv \Lambda_\mathrm{shuttle}^{0.12}\,.
\end{align}

The distance travelled per unit cell by line service, $\Delta_\textrm{train}$, remains constant throughout our study because trains operate at a constant frequency $\mu$, mathematically,
\begin{equation}
    \Delta_\textrm{train} = 4\cdot\mu\cdot\ell\,.
\end{equation}

$\Delta_\textrm{MIV}$ is the total distance requested by users per unit time, 
\begin{equation}
\label{eq:delta_miv}
    \Delta_\textrm{MIV} = \nu E \ell^{2}D\,.
\end{equation}

Analytically, Eq.~\ref{eq:emission_word} can then be written as
\begin{equation}\label{eq:emission_math}
\begin{aligned}
    \emissions=  \underbrace{\eta^{-1}\left(\dldctilde\funi + 2\beta\tilde{\ell} \fbi \right)\cdot\frac{\shuttleEnergy}{\carEnergy}}_\textrm{shuttles}   + \underbrace{\frac{4\tilde{\mu}}{\Lambda \tilde{\ell}}\cdot\frac{\trainEnergy}{\carEnergy}\,}_\textrm{train}.
\end{aligned}
\end{equation}
We consider electric light rails with a maximum seating-capacity $k=100$ and $\trainEnergy = 9.72\, \unit{kN}$ \citep{TREMOD} for the line service. For MIV we consider Diesel cars with $\carEnergy=2.47\, \unit{kN}$ \citep{viz2021}.
For the shuttles we choose Mercedes Sprinter ($8.8$ liters of Diesel per $100\, \unit{km}$ \citep{sprinter_energy}), resulting in $\shuttleEnergy = 3.28\, \unit{kN}$.

In order to compute the ratio above in ~Eq.~\ref{eq:emission_word} for a simulated scenario, we perform independent simulations for MIV and bi-modal transit with identical user requests. The denominator in ~Eq.~\ref{eq:emission_word} is obtained from MIV simulations by multiplying the total driven distance by $\carEnergy$. The numerator in ~Eq.~\ref{eq:emission_word} is obtained from bi-modal simulations by multiplying mode-specific total driven distance with the respective vehicle-specific energy consumption per unit distance.

%
%
%

\paragraph{Traffic}
\label{subsec:traffic_definition}
Road traffic is a source of local noise and air pollution and occupies significant shares of urban space. Bi-modal transit aims at the reduction of road traffic by utilizing line services for trips over larger distances.

To obtain a quantitative estimate, we define as traffic volume for the two used cases, $\tilde{\Delta}$, cumulative distance driven by shuttles, $\deltashuttle$ normalized with respect to the equivalent of total MIV distance requested, $\deltamiv$ (Eq.~\ref{eq:delta_miv}), or, equivalently, the relative number of driving shuttles as compared to MIV.
\begin{equation}
\label{eq:traffic}
\tilde{\Delta} \equiv \frac{\Delta_\textrm{shuttle}}{\Delta_\textrm{MIV}}=\etabi^{-1}\cdot \, (1+F) \,\dshuttletilde\,,
\end{equation}
with pooling efficiency, $\eta$, bimodal fraction, $\fbi$, and average requested distance for trips by shuttles involved in bi-modal transit, $\tilde{D}_\textrm{shuttle}$.

\section{Results}
\label{sec:results}
Below in Subsec.~\ref{subsec:drrp_performance}, we first present how DRRP performs with a bi-modal system when the bi-modal transit system is used by $10\%,$ and $1\%$ population. Then in Subsecs.~\ref{subsec:performance} , and \ref{subsec:pareto}, we describe the overall performance of the bi-modal transit system for the two used cases above. We conclude the results section with an analysis of the potential reduction in traffic volume.

\subsection{DRRP performance}
\label{subsec:drrp_performance}
\begin{figure*}
    \centering
   \includegraphics[width=0.8\textwidth]{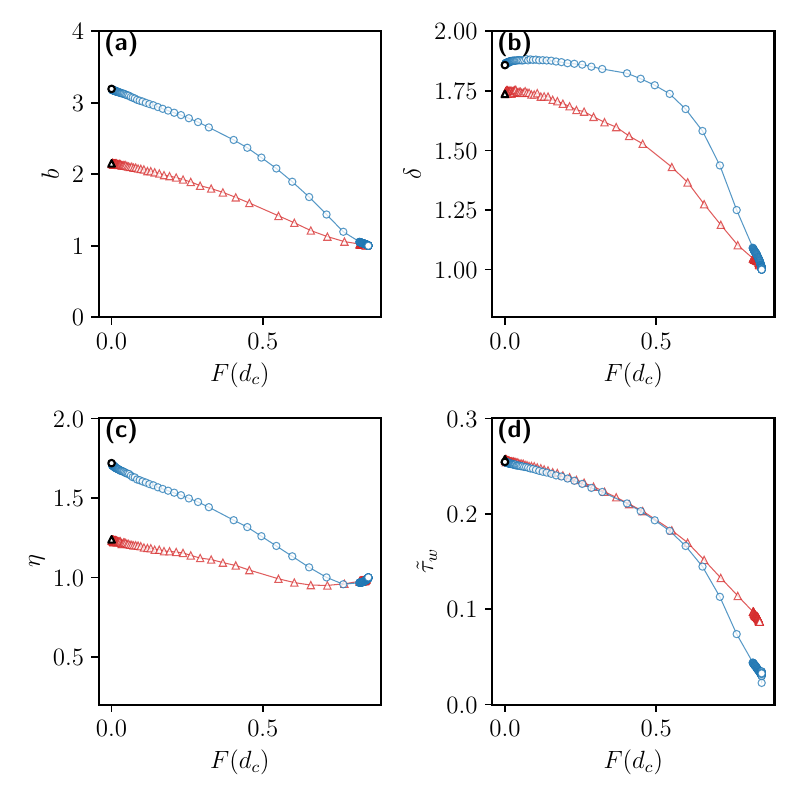}
     \caption{\textbf{DRRP performance statistics:} DRRP/Shuttle performance parameters are plotted against bi-modal fraction $\fbi$. Blue circles and red triangles represent $10\%,$ and $1\%$ population of Greater Berlin, respectively. 
     Black circle, and triangle represent uni-modal transport (shuttles only) \textbf{(a)} Mean DRRP occupancy for non-standing vehicles, $b$. \textbf{(b)} Mean detour, $\delta$, for shuttle users. 
     \textbf{(c)} Mean DRRP pooling efficiency $ \eta \equiv b/\delta$.
     \textbf{(d)} Mean waiting time, $\tilde \tau_w$, for shuttles normalized with $t_{0}$.}
     \label{fig:DRRP_stats}
\end{figure*}

\paragraph{Occupancy} In Fig.~\ref{fig:DRRP_stats}a, we show the mean DRRP occupancy, $b$, averaged over non-empty driving vehicles against the bi-modal fraction, $F$, for $10\%$ and $1\%$ use case. We observe that the shuttles have a higher mean occupancy for larger use case, i.e., larger demand because of the greater possibility of pooling. Mean occupancy decreases with the involvement of trains (increasing $F$). This is because shuttle trips are shortened causing passengers to spend less time in shuttles during their transit. The black symbols represent mean occupancy for a uni-modal (shuttles only) scenario. We observe that the mean occupancy is larger for uni-modal scenarios.

\paragraph{Detours} In Fig.~\ref{fig:DRRP_stats}b, we observe that higher use case, i.e., $10\%$ population has larger detours. Detour $\delta$ and $b$ trend observed above shows a well-known trade-off between detour and pooling for DRRP, i.e., desirable pooling necessitates undesirable detours for passengers \citep{herminghaus_mean_2019}. We observe that the detours decrease with the involvement of line services, this can be attributed to the 'common stop effect', a phenomenon observed in the previous study \citep{intermediate}. With greater involvement of line services, more passengers are picked up and dropped of at the same train station, thereby reducing detours.

\paragraph{Pooling efficiency} The pooling efficiency which is defined as the ratio between mean occupancy and mean detour is shown in Fig.~\ref{fig:DRRP_stats}c. We observe that the pooling efficiency, $\eta$, is higher for larger demand, i.e., $10\%$ use case as also reported in previous studies \citep{muehle2022,intermediate}. This suggests that the larger use of shuttles or bi-modal service will be favorable for pooling efficiency. We observe that the involvement of line service reduces DRRP pooling efficiency because user requests are diverted toward the line which shortens the average distance a passenger travels on the shuttle during the entire journey. 

\paragraph{Waiting time} In Fig.~\ref{fig:DRRP_stats}d, we study the mean waiting time for shuttle-borne trips. We plot the mean waiting time normalized with the average trip duration, $t_{0}$, against the bi-modal fraction, $F$. We observe that larger demand, that is, the use case of ten percent has larger waiting times because shuttles are busier. We also observe that the involvement of line services decreases the waiting time for shuttle trips because of the 'common stop effect' and a lower share of distance traveled in shuttles. 

The main messages from Fig.~\ref{fig:DRRP_stats} are: 1) Shuttles become more efficient with demand, while user experience suffers due to larger detours and waiting times, as also found in the previous study \citep{intermediate}. 2) The involvement of line services makes the shuttle trips more convenient for users by reducing the waiting time and detours.

\subsection{Overall energy consumption and service quality of bi-modal transit}
\label{subsec:performance}
Now, we will analyze the overall objectives, i.e., energy consumption (Eq.~\ref{eq:emission_math}) and service quality (Eq.~\ref{eq:quality_math}) of the transportation system for the two use cases.

\begin{figure*}
    \centering
    \begin{subfigure}[b]{\textwidth}
    \centering
    \includegraphics{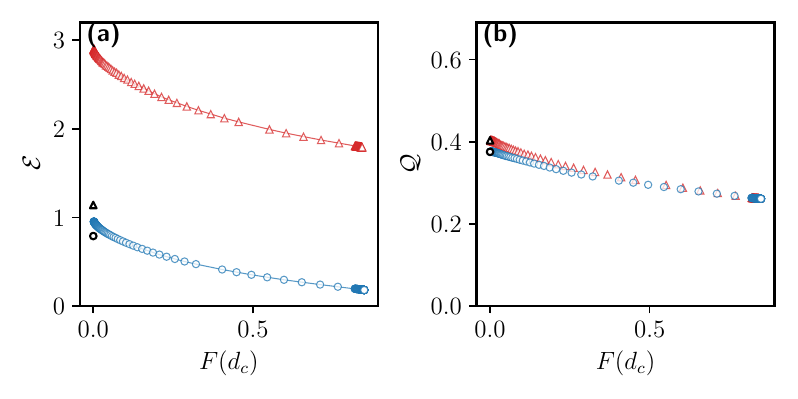}
    \end{subfigure}
    \begin{subfigure}[b]{\textwidth}
    \centering
    \includegraphics{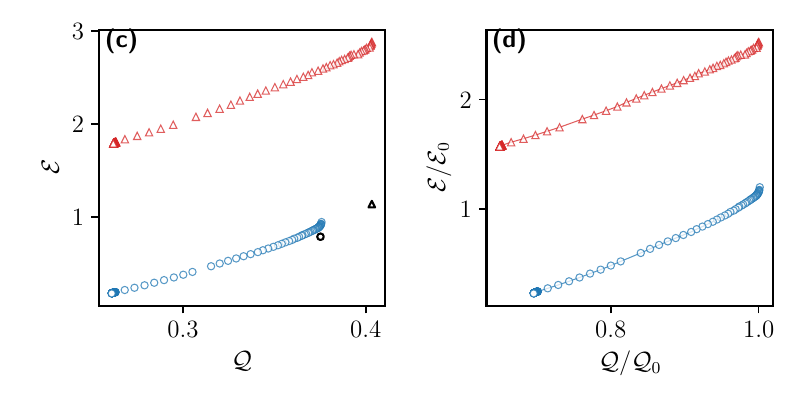}
    \end{subfigure}
     \caption{\textbf{Overall performance of bi-modal transit: } Blue circles and red triangles represent data for $10\%$, and $1\%$ greater Berlin population respectively.  
     \textbf{(b)} Quality, $\quality$, as a function of $\fbi$.
     \textbf{(c)} Pareto fronts of energy consumption, $\emissions$, vs. service quality, $\quality$ determined from the data shown in (a), (b). Data not part of Pareto fronts is not shown. 
     The black circle and triangle represent uni-modal transport (shuttles-only) data for $10\%$, and $1\%$ greater Berlin population respectively. \textbf{(d)} Pareto fronts as in (c), but normalized with respect to the performance, $(\quality_{0},\emissions_{0})$, of the uni-modal system (shuttles only).}
     \label{fig:emission_quality}
\end{figure*}

\begin{figure*}
    \centering
    \includegraphics[width=0.8\textwidth]{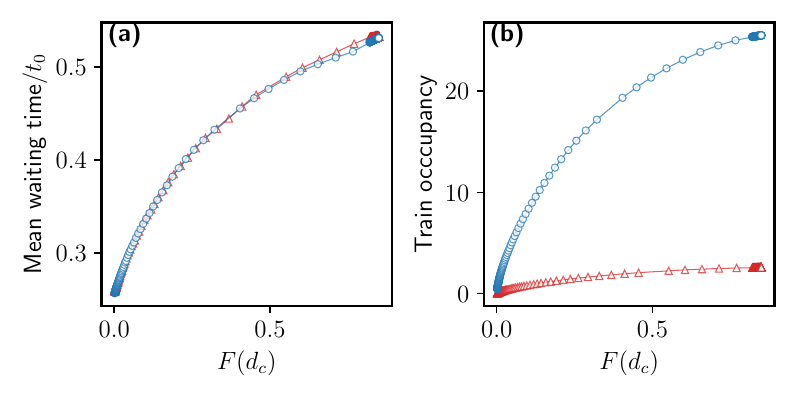}
    \caption{\textbf{Mean waiting time and Train occupancy:} Blue circles and red triangles represent $10\%$ and $1\%$ population of Greater Berlin, respectively. \textbf{(a)} Mean waiting times normalized with $t_{0}$. \textbf{(b)} Mean train occupancy.}
    \label{fig:pt_stats}
\end{figure*}

\begin{figure}
    \centering
    \includegraphics{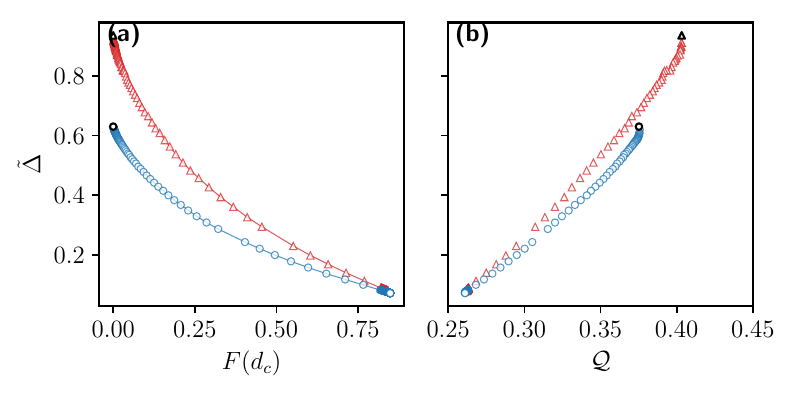}
    \caption{\textbf{Traffic volume.}
    Blue circles and red triangles represent $10\%$, and $1\%$ greater Berlin population respectively. \textbf{(a)}Relative bi-modal traffic in simulations, $\traffic$, as defined in Eq.~\ref{eq:traffic}, as a function of the bi-modal fraction, $F$. 
    Black circle and triangle represent uni-modal transport (shuttles-only) data for $10\%$, and $1\%$ greater Berlin population respectively. \textbf{(b)} Relative bi-modal traffic in simulations, $\traffic$, determined along the Pareto fronts in Fig.~\ref{fig:emission_quality}, against corresponding service quality $\quality$.
    }
    \label{fig:traffic}
\end{figure}

\paragraph{Energy consumption} In Fig.~\ref{fig:emission_quality}a, relative energy consumption, $\emissions$, is plotted as a function of bi-modal fraction, $F$, for the two use cases discussed above. We observe a general trend of decreasing energy consumption with the involvement of line services. This is because $\deltashuttletilde$ decreases with the involvement of line services (see Fig.~\ref{fig:traffic}a) while $\tilde{\Delta}_\textrm{train}$ is constant due to a fixed schedule of trains in the simulations, thus reducing the total energy consumption by the bi-modal transportation system. We also observe that energy consumption is reduced with increasing demand, i.e., for a higher used case. This is evident from Eq.~\ref{eq:emission_math}. Note that the emission curves for bi-modal transportation start above the emissions for uni-modal scenarios (black symbols). This is because in our bi-modal simulations, trains are always running at fixed schedules. For low bi-modal fraction, $\fbi$, trains are underutilized because they operate at low occupancies (see Fig.~\ref{fig:pt_stats}b). We observe that for $10\%$ use case, the energy consumption can easily drop below $20\%$, however, for a $1\%$ use case, it's not advisable to use bi-modal transportation, suggesting that a larger adoption of bi-modal transportation can significantly reduce the energy consumption.

\paragraph{Quality} In Fig.~\ref{fig:emission_quality}b, the overall quality of the system is plotted against the bi-modal fraction, $F$, for the two use cases. We observe that the demand doesn't significantly impact the overall service quality and  service quality decreases with the involvement of the line services. Large waiting times (see Fig.~\ref{fig:pt_stats}a) contribute to degrading quality with the bi-modal fraction, $\fbi$. This suggests that user quality can be improved by increasing the train frequency and adapting the train capacity, $k$, depending on the demand. We see in Fig.~\ref{fig:pt_stats}b that trains are barely full.

\subsection{Pareto optimization}
\label{subsec:pareto}
A tuple of parameter values, in our case $(\emissions,\quality)$, is called Pareto-optimal if none of the parameters (or objectives) can be further optimized without compromising on at least one of the others.
The set of all such tuples of parameters is called the Pareto front \citep{debreu1959,greenwald1986,magill2002}.
We now apply this concept to our results, keeping in mind that we aim at maximum service quality at minimum energy consumption.
Hence, in diagrams spanned by  $\quality$ as the abscissa and $\emissions$ as the ordinate, system operation as far as possible to the lower right is desirable. 

In order to study the overall performance of bi-modal transportation in real scenarios and the impact of user adoption, we explore two distinct scenarios: one where $1\%$ of the total population utilizes bi-modal transportation, and another where the adoption rate increases to $10\%$ of the total population. In each case, we vary $\dcut$ to obtain the Pareto fronts.

In Fig.~\ref{fig:emission_quality}c, we show the Pareto fronts obtained for data in Figs.~\ref{fig:emission_quality}a,b. We observe that the energy consumption can go below $20\%$ for a service quality of around $0.25$ for the $10\%$ use case. The black circle represents the uni-modal data for the $10\%$ use case. We observe that bi-modal transportation can significantly reduce emissions as compared to uni-modal (shuttles only) case with some compromise on service quality. This is clear from Fig.~\ref{fig:emission_quality}d, where we plot the Pareto-optimal $\emissions$ and $\quality$ normalized with uni-modal $\emissions_{0}$ and $\quality_{0}$ respectively.

For the $1\%$ use case, it is not advisable to deploy bi-modal transportation at all because requests are better served by uni-modal transportation (black triangle) both in terms of energy consumption and service quality. 

\paragraph{Traffic volume}
\label{subsec:traffic_results}
In Fig.~\ref{fig:traffic}a, we plot the total relative traffic volume, $\tilde{\Delta}$, described above in subsec~\ref{subsec:traffic_definition} on the vertical axis as a function of bi-modal fraction, $\fbi$, for the two used cases. We observe a trend of decreasing bi-modal traffic with the involvement of line services, i.e., with increasing $\fbi$. Also, $\tilde{\Delta}$ decreases when the user adoption goes from $1\%$ to $10\%$ because the shuttles become more efficient due to the possibility of larger pooling (see also Fig.~\ref{fig:DRRP_stats}c). 

In Fig.~\ref{fig:traffic}b, we plot the total relative traffic volume, $\tilde{\Delta}$, for simulations, determined along the Pareto fronts in Fig.~\ref{fig:emission_quality}c, against corresponding service quality, $\quality$. The traffic volume for uni-modal (shuttles-only) scenarios is plotted with black symbols. We observe that the relative traffic volume for uni-modal scenarios decreases with demand due to increased pooling efficiency, $\eta$ (see Fig.~\ref{fig:DRRP_stats}c). The relative traffic volume for uni-modal (shuttles only) scenario for $1\%$ use case is not significantly less than that of MIV because of low pooling efficiency, $\eta$ (see Fig.~\ref{fig:DRRP_stats}c). The uni-modal traffic volume for the $10\%$ use case is around $60\%$ of MIV traffic. This suggests that the relative uni-modal traffic volume can be further reduced if more people adopt ride pooling. 

Bi-modal public transportation allows for further reduction in relative traffic volume Below the uni-modal scenario, albeit, at a lower service quality. The bi-modal traffic can go as low as $15\%$ for $10\%$ user adoption. 

\subsection{Discussion}
Our investigation had two primary goals. First to study the feasibility of bi-modal demand-responsive public transportation in Berlin and Brandenburg with the existing rail network. Second, to study the impact of user adoption of public transit on the overall performance of bi-modal demand-responsive public transportation in Berlin and Brandenburg. 

Our study suggests that bi-modal demand-responsive transportation can be deployed in Berlin and Brandenburg with the existing rail network of public transportation. We find that the overall performance of bi-modal transportation improves with higher user adoption. While $10\%$ user adoption can significantly reduce emissions and vehicular traffic, it is not advisable to deploy bi-modal transit with existing rail network and train schedules if the user adoption is $1\%$. This suggests that the overall performance of the bi-modal transit can be further improved by devising strategies to attract users towards bi-modal transportation. 

\bibliographystyle{elsarticle-num-names}
\bibliography{main}

\end{document}